# Correlating atom probe tomography with X-Ray and electron spectroscopies to understand microstructure-activity relationships in electrocatalysts

Baptiste Gault[1,2,*], Kevin Schweinar[1], Siyuan Zhang[1], Leopold Lahn[1,3,4], Christina Scheu[1], Se-Ho Kim[1], Olga Kasian[1,3,4]

[1] Max-Planck-Institut für Eisenforschung, Düsseldorf, Germany.

[2] Department of Materials, Royal School of Mines, Imperial College London, London, UK

[3] Helmholtz Institut Erlangen-Nürnberg, Helmholtz-Zentrum Berlin GmbH, Hahn-Meitner-Platz 1, 14 109 Berlin

[4] Department of Materials Science and Engineering, Friedrich-Alexander-Universität Erlangen-Nürnberg, 91 058 Erlangen, Germany

[*] Corr. Author: b.gault@mpie.de

## Abstract

The search for a new energy paradigm with net-zero carbon emissions requires new technologies for energy generation and storage that are at the crossroad between engineering, chemistry, physics, surface and materials sciences. To keep pushing the inherent boundaries of device performance and lifetime, we need to step away from a *cook-and-look* approach and aim to establish the scientific ground to guide the design of new materials. This requires strong efforts in establishing bridges between microscopy and spectroscopy techniques, across multiple scales. Here, we discuss how the complementarities of X-ray- and electron-based spectroscopies and atom probe tomography can be exploited in the study of surfaces and sub-surfaces to understand structure-property relationships in electrocatalysts.

# 1 Introduction

The technologies necessary for the electrification of transportation, the generation of green hydrogen at scale[1] for transport and manufacturing[2], and the grid-scale energy storage to accommodate the intermittency of renewable electricity generation[3,4], all require new materials solutions[5]. The performance and service life time of these materials are inherently related to their microstructure, chemistry, physics and interaction with their environment. Understanding the subtle processing-structure-property relationships is key to guide the design of future generations of materials. Yet, the variety of scales involved, from individual atoms to nano- to micro- to millimeter-scales, and nature of the information – i.e. surface atomic arrangements, crystalline structure, composition and chemistry – is simply too broad for a single technique to provide all the necessary insights. Correlative approaches must be hence developed to build upon the strength of each individual technique. Another challenge is that a material's chemical (re)activity can be underpinned by metastable species at the surface of the electrodes and their analyses can only be performed by using in situ or in operando techniques.

Atom probe tomography (APT) provides three-dimensional compositional mapping of materials with sub-nanometer resolution[6] and as such is poised to provide extremely valuable insights. APT has progressively increased in prominence in the arsenal of characterization techniques for bulk metallic materials[7–9], increasingly for ceramics and semiconductors [10,11], phase change materials[12], and with recent forays into nanostructures[13,14] including nanoparticles[15–19] and e.g. 2D-materials[20,21] with potential application in catalysis, as recently reviewed by Baroo et al. [22]. There is a compatibility in scales between specimens for APT, shaped as sharp needles with a near-spherical cap at their end, and nanoparticles used across many catalysis applications, i.e. 10–200 nm in diameter. Historically, this had enabled field-electron emission microscopy

(FEEM/FEM) to be extensively used to study surfaces in reaction conditions[23]. The reacted surfaces were then imagined by field-ion microscopy (FIM) in search for morphological or atomic-scale topographical modifications. Early implementation of atom probes were then sometimes used to measure changes in the surface chemistry, however they were limited to one-dimensional depth-profiling of the composition or to study a single set of reacted planes[23].

The analysis of catalysts and electrocatalysts is however an area in which deploying the full potential of APT could be transformational. However, despite outstanding works performed in a limited number of groups [24–27], APT has not seen a fast spread in surface sciences studies, in part because of difficulties in specimen preparation but also a lack of direct structural information available along with the chemical and bonding state of surface species. These are accessible through X-ray photoelectron spectroscopy (XPS), one of the key techniques used in surface sciences. XPS allows access to the bonding state within the first few atomic layers of the surface. For its compatibility in scale, APT is more often correlated with (scanning) transmission electron microscopy, in which the bonding state can be obtained from electron energy loss spectroscopy ((S)TEM- EELS)[28,29]. Preliminary work on thin films[30,31] demonstrated the importance of combining XPS, STEM-EELS and APT to get a holistic understanding of the microstructural origin or the materials performance and degradation mechanisms.

In this article, following a brief overview of the working principles of these techniques to help the reader get to speed, we discuss the complementarities of these techniques and showcase examples from the recent literature to provide a perspective on the strength of the correlative approaches.

## 2 Working principles & techniques' complementarity

## 2.1 Working principles

We briefly introduce the working principles of the techniques discussed below for non-experts. This should in any case not be seen as an exhaustive overview of these techniques, merely a segue enabling to position the perspective on the correlative approaches.

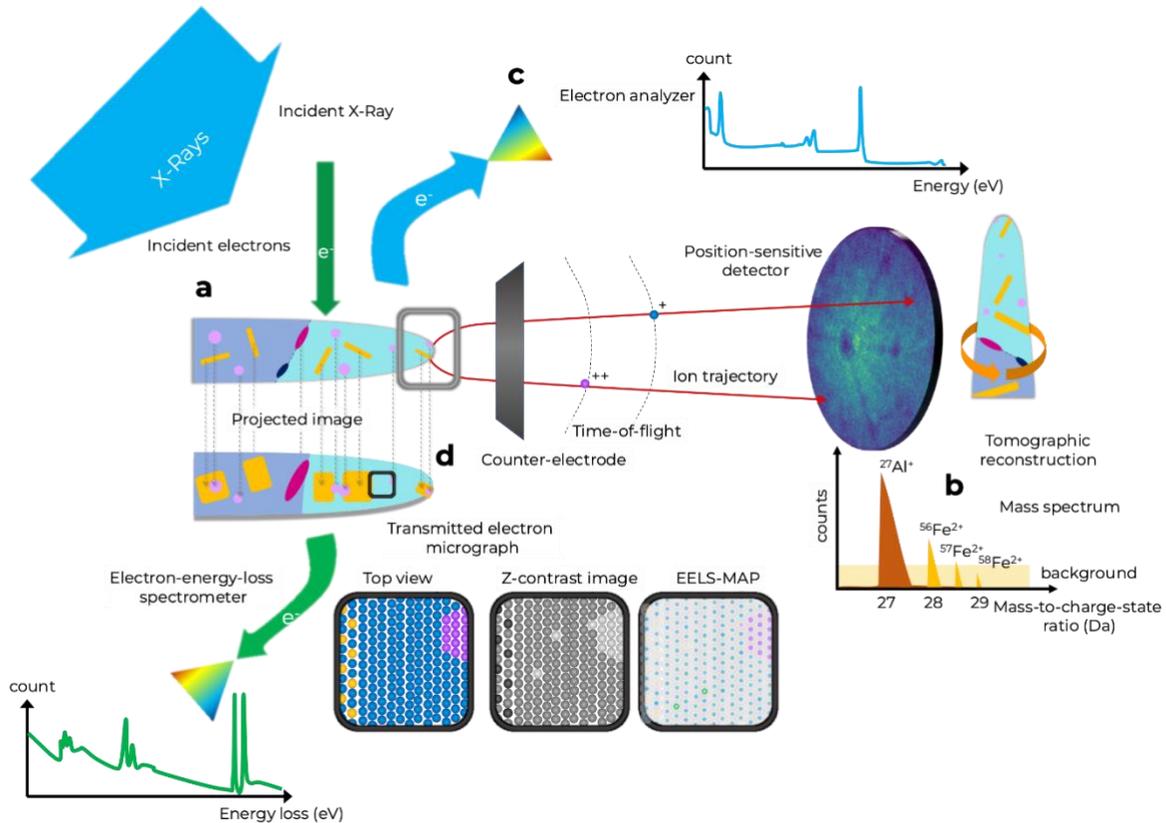

Figure 1: (a) cartoon-view of the specimen in an atom probe; (b) main results from APT; (c) schematic view of XPS performed on a similar sample; (d) projected view of the APT specimen as obtained from (S)TEM with an EELS spectrometer and possible high-resolution images and chemical maps.

### 2.1.1 Atom probe tomography

In APT, as schematically depicted in Figure 1a, the atoms leave the surface of a needle-shaped specimen successively in the form of ions under the influence of an intense electric field. Following field evaporation, these ions fly along well-defined trajectories[32,33] and are collected by a single-particle detector that records the impact position[34]. The field evaporation is time-controlled by using either fast voltage[35] or laser pulses[36,37] superimposed to a direct current (DC) high-voltage, thereby enabling the identification of the elemental nature of each ion by time-of-flight mass spectrometry[38]. The results from an APT experiment take the form of a mass spectrum, i.e. a histogram of the number of ions detected at a certain mass-to-charge ratio, whereas the detector impact positions are used to build a point cloud providing the atomic distribution in three-dimensions[39], as summarized in Figure 1b.

2.1.2 X-ray photo-emission and electron energy loss spectroscopies

In XPS, a micron-to-millimeter sized beam of X-rays is focused onto the sample's surface. Upon penetration, some electrons from the material itself absorb photons from the incoming beam and get ejected from the material – i.e. the photo-electric effect. By measuring the kinetic energy of the emitted electron, and knowing the energy of the incoming photon, the bonding energy can be readily determined[40]. This is schematically summarized in Figure 1c. XPS hence provides a precise account of the chemistry of the first few atomic layers at the specimen's surface, averaged over microns-to-millimeters of the surface depending on the setup that is used[40]. There are ongoing efforts to perform spatially-resolved XPS experiments with a lateral resolution in the range of tens-to hundreds of nanometers[41]. A clear strength of XPS is the possibility to be used *in-operando,* i.e. at temperatures and gas pressures that can mimic service conditions, to provide a precise account of the evolution of the surface chemistry over the course of reactions[42].

STEM-EELS can provide similar information: as the incoming electrons travel through the thin specimen, inelastic scattering causes the loss of amounts of energy that can be related to plasmons or to the ionization energy of specific species. Analyzing the spectrum of the kinetic energy of the electrons coming out of the specimen allows for determining the specimen's composition and bonding state of the different species present. As the electron-beam in STEM is tightly focused and scanned across the specimen's surface, STEM-EELS enables mapping of the species potentially within each individual atomic-column along the pathway of the electron beam [43,44], as schematically shown in Figure 1d. There are also ongoing efforts to enable *in-situ* or quasi *in-operando* observations by STEM-EELS[45] to help better understand materials processes in conditions mimicking service.

## 2.2 Complementarities & challenges

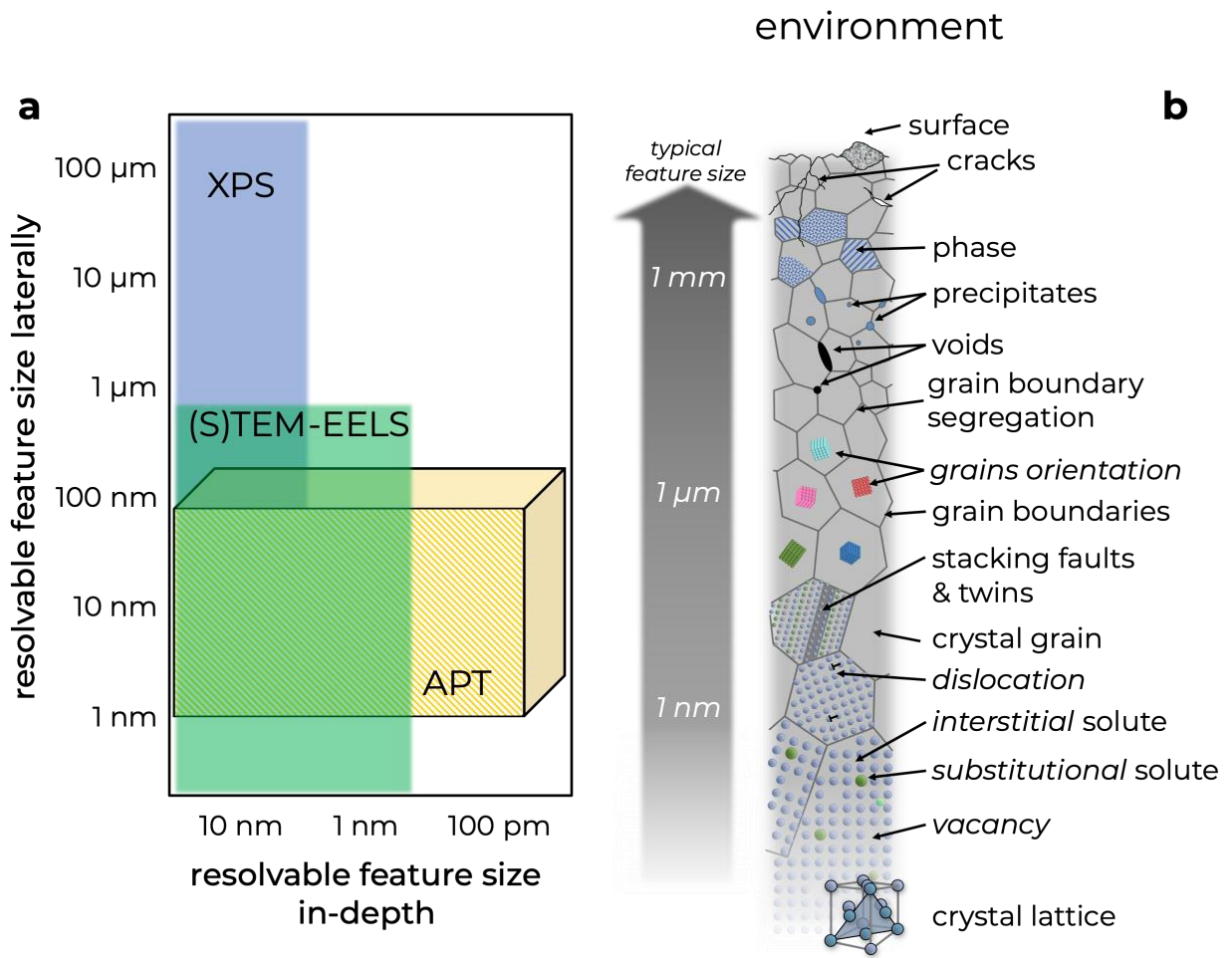

Figure 2: (a) summary view of the resolvable feature size for APT, STEM-EELS and XPS. (b) schematic view of the main microstructural features and their typical size to provide a perspective on what can be addressed by the various techniques discussed herein.

First, there is a complementarity in the scale of the features that can be analyzed by the different techniques, and a compatibility in the spatial resolutions. The spatial resolution of each of these techniques is anisotropic and depends on an array of parameters. What matters more is not the absolute value of the spatial resolution but the size of the smaller resolvable feature both laterally and in depth. For both XPS and STEM-EELS, the spatial resolution not only depends on intrinsic instrumental parameters – the spot size of the illuminating beam –, but also extrinsic factors – for instance the nature of the specimen itself. Indeed, in XPS, the penetration depth is dependent on the composition itself, typically below 10 nm, whereas the lateral extent depends on the size of the focused X-ray beam. In STEM-EELS, the signal collected is integrated through the thickness of the thin specimen (20–100 nm) and depends on the lateral extent of the focused electron beam, and its potential spread from scattering going through the specimen[46]. The actual spatial extent of the probed volume of the material can hence be difficult to ascertain. This is where the inherent strength of APT can be best used. In APT, the spatial resolution or size of the smaller resolvable microstructural feature of interest depends on its composition [47,48], but can be assumed to be below 1 nm in all three dimensions[49] but is typically an order of magnitude better in depth[47,50], as summarized in Figure 2a.

Now, let us consider common microstructural features, for instance stacking faults, dislocations, grain boundaries, and secondary phases, along with their respective dimensions, as shown in Figure 2b. The signal in the recorded spectra from XPS or STEM-EELS can be extremely complex as it is a convolution of the signals originating from the features within the probed volume. Fitting approaches are hence used to decompose the signals originating from the different bulk or surface features averaged over the volume probed by the electron or X-ray beam[29,44,51]. This can lead to components and phases with a low volume fraction to be missed or overlooked, yet their influence on the materials properties may remain substantial.

This is where the correlation with APT can become a game changer: APT provides compositional mapping in three-dimensions within the reconstructed volume, with a compositional sensitivity that can be in the range of parts-per-million[52], and individual features can be interrogated separately after data reconstruction and segmentation. This information can then be used to better understand the origins of the signal in the XPS or EELS spectrum. In turn, the bonding state remains inaccessible to APT. This is a first critical aspect of the complementarity with XPS and STEM-EELS, as the bonding state is required to understand the chemical activity or reactivity of a compound.

APT has a higher chemical sensitivity[52] than these spectroscopic techniques, especially for light elements, e.g. H[53,54], C, and N[55]. Yet the compositional accuracy strongly depend on the local intensity of the electric field[56–58] during the analysis. These issues can make the analysis of certain materials or features challenging, and the complementary insights into the composition of some interfaces for instance, that can be gleaned from STEM-EELS, can help support [59,60] the validity of some measurements.

Finally, the analysis of a sample's outermost surface requires dedicated strategies to avoid damage and contamination during specimen preparation and transport. For APT and STEM-EELS, specimens are typically prepared using a focused ion beam (FIB)[61,62], in which the energetic incoming ions get implanted, can cause amorphization and can push surface atoms to penetrate inside the sub-surface region[63–65]. The use of cryogenic temperatures during the preparation can help alleviate some issues [66,67], but not all. A thin metallic coating can be used to protect the very surface, yet it often involves transporting the sample through ambient air, which can have an influence on the surface species themselves. There are efforts to develop approaches to transport samples under protective environments or (ultra-) high vacuum conditions, and possibly under cryogenic conditions [68–71]. The use of transport suitcases has been more common in surface sciences, and most XPS instruments are equipped with an ion-gun to clean the specimen's surface by sputtering, and remove the first few atomic layers.

# 3 Workflows and applications

Challenges arise in the development of appropriate workflows enabling analysis of the same sample to correlate the signals from the same microstructural features. Schweinar et al. proposed to use thin-films deposited on the commercial flat-top coupons[72] used as support for APT specimen preparation[73,74], as shown in Figure 3a. They performed spatially-resolved XPS directly onto the film before and after thermal oxidation, Figure 3b–c, revealing the change in the oxidation state of Ir and Ru atoms. The compositional maps appear homogeneous across the 3 µm-wide disc at the top of the microtip. Following this analysis, the sample's surface was protected by a thin layer of sputter-deposited Cr, and subsequently sharpened into an APT specimen by FIB [75] and analyzed by APT. The corresponding APT reconstruction is shown in Figure 3d. The analysis starts from the Cr capping layer, goes through the oxide layer, and terminates into the metallic film. Interestingly, by looking at a region-of-interest across the x-y section of the dataset in the oxidised region, Ru is not spread across the material but rather agglomerated along grain boundaries within the thin film, as readily visible in Figure 3e. APT allows for quantification of the Ru segregation, determined as up to nearly 10 at%, Figure 3f, accompanied by a depletion of Ir. This observation is likely related to the reduction in the free energy of the grain boundary associated to the segregation of Ru prior to oxidation.

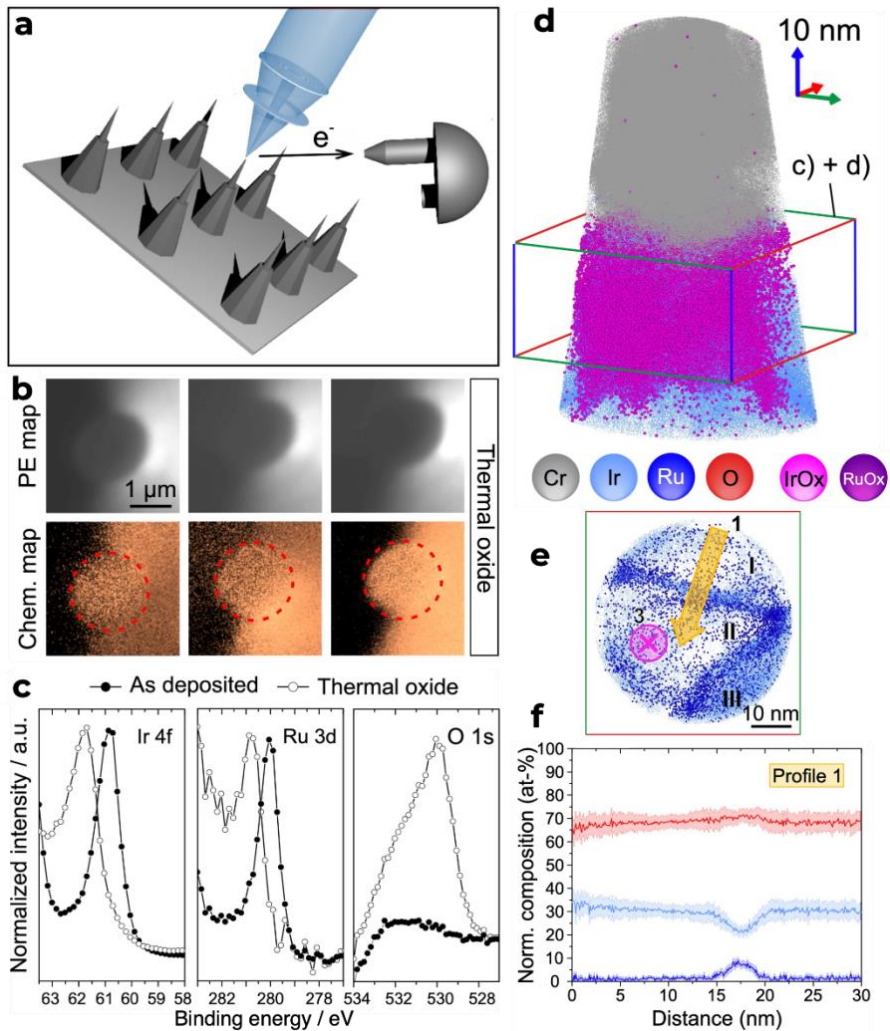

Figure 3: Composite image of the workflows introduced by Schweinar et al. in Ref. [73,74] (reproduced and modified under a CC-BY licence). (a) Schematic of the analysis by spatially-resolved XPS of a thin Ir-Ru film deposited on flat-top commercial microtip coupons, normally used as support for lift-out specimen preparation for APT. (b) Photo-electron map and seemingly homogeneous compositional map obtained from the film after thermal oxidation at 600 °C for 5 h in air. (c) Comparison of spectra before and after oxidation showing the metallic state and oxidized state, respectively. (d) reconstructed APT dataset following deposition of a Cr capping layer and sharpening. (e) Cross-section through the APT reconstruction showing segregation of Ru to grain boundaries within the oxidised film. (f) Composition profile calculated in a region of interest positioned across the grain boundary enabling quantification of the Ru (blue) segregation.

This exemplifies the kind of insights that APT can bring – XPS could only detect a single oxide phase, whereas a richer microstructure is resolved by APT. The local changes in composition at grain boundaries, combined with a different local atomic organization will change the catalytic activity of the grain boundaries, as other studies have also shown with the aid of STEM-EELS [30]. If APT can be performed on a thin film suitable for XPS, the opposite was also proposed by Balakrishnan et al.[76], in Figure 4a. They performed XPS first on APT specimens of Ir, which had been used as an electrocatalyst for the oxygen evolution reaction. Following XPS, the specimens were then analyzed by APT to reveal the local compositional evolution of the different sets of atomic planes that form the end surface of the needle.

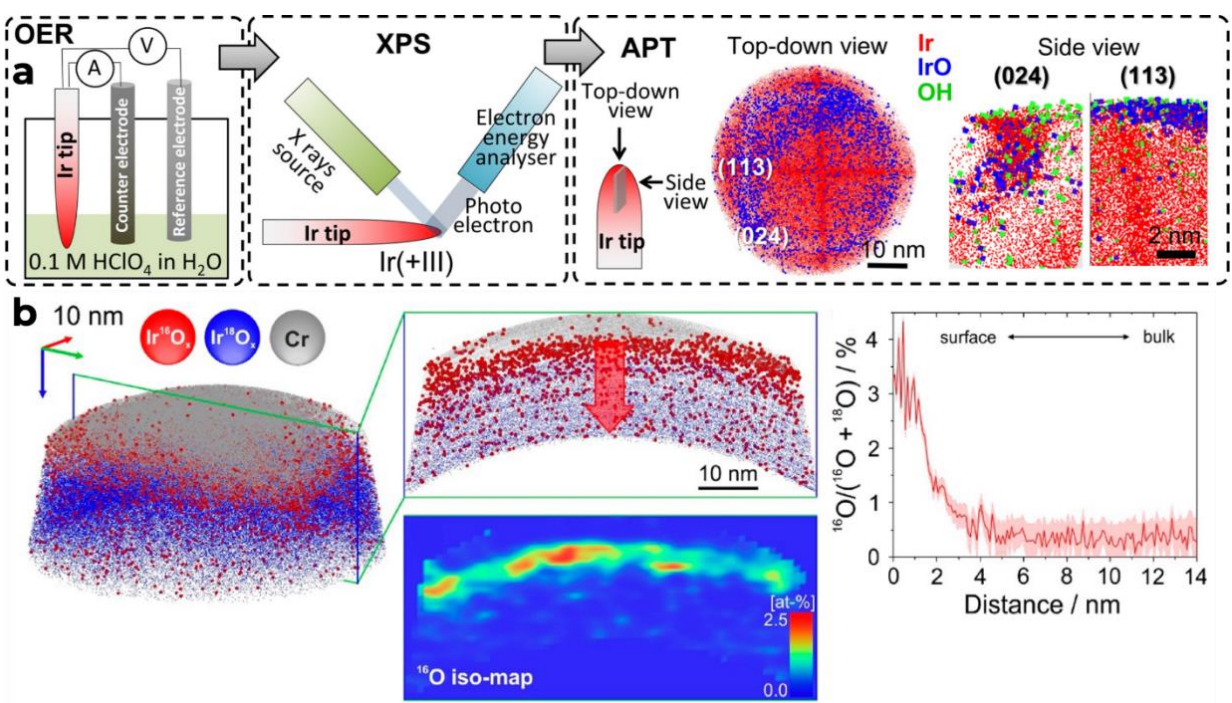

Figure 4: (a) workflow for the analysis of APT specimens used as electrocatalysts by XPS (from [76] reproduced under a CC-BY licence). (b) summary of the oxygen isotope exchange experiment on iridium oxide during the oxygen evolution reaction (from [77] reproduced under a CC-BY licence).

Both of these approaches have advantages and drawbacks. For instance, the APT specimen's shape and size can make direct electrochemical or XPS measurements not necessarily straightforward, making the latter approach not always suitable, particularly when site-specific analysis is necessary. Yet the need to shape the specimen with a FIB after the reaction typically requires transporting the specimen through air, and the energetic ion beam can induce damage to the reacted sample's surface. Transport under protective atmospheres can be possible by using dedicated infrastructure [68] but it is not yet routinely available.

In any case, these studies showcased the spatial sensitivity of APT and the importance of the precise assessment of the species on the surface and in the near-surface regions. The depth probed by XPS is often not known precisely, as the penetration of the X-rays depends on the local composition and structure. In contrast, APT can provides direct insights on features on the surface, provided that appropriate protection of the surface is used prior to analysis, and the spatial resolution of APT in the depth is extremely high [48,50]. There remains some uncertainties estimated to be 10% or more [78] in the depth dimension of the APT reconstruction [79]. However, calibration of the spatial reconstruction parameters is possible either by using the partial structural information[80] from within the data or through the use of TEM[81]. This leads to an extreme spatial resolving power of APT which was, for instance, combined with isotope labelling to investigate the exchange of oxygen between the solution and the lattice of iridium oxide during the oxygen evolution reaction [77], as summarized in Figure 4b.

Most showcase studies so far have focused on the analyses of electrocatalysts and electro-photo-catalysts that find application in hydrogen energy cycle, especially the analysis of Ir metal in the form of thin films[30], metal wire[76] and mixed Ir-Ru alloys[73], but also an array of oxides [13,31,82–84], in particular to better understand the microstructural origins of their activity [31] or degradation in service [83]. In principle, similar approaches could be deployed in the future to a broad range of electro(photo)catalysts.

# 4  Conclusions

APT arises from field-ion microscopy, and both were originally surface science techniques, yet progress in APT made the community progressively turn its attention towards bulk analyses. Yet APT with its intrinsic capacity to map the composition or a material with a resolution better than a nanometer in three-dimensions has great potential for complementing XPS that can probe the chemistry of a surface, and STEM-EELS which provides compositional and chemical mapping through the thickness of a thin sample. Information from APT can guide the fitting of the electron energy spectra and facilitate data interpretation, as well as support investigation into processing-microstructure-property relationships that are necessary to design new electrocatalysts and electro-photo-catalysts. The difference in scale can be difficult to reconcile but the development of dedicated workflows has already enabled progress that will continue in the decade to come.

# Acknowledgements

We thank Uwe Tezins, Christian Broß and Andreas Sturm for their support to the FIB and APT facilities at MPIE. We are grateful for the financial support from the BMBF via the project UGSLIT and the Max-Planck Gesellschaft via the Laplace project. S.-H.K. and B.G. acknowledge financial support from the ERC-CoG-SHINE-771602. K.S. is grateful for funding from the IMPRS-SurMat.

On behalf of all authors, the corresponding author states that there is no conflict of interest.

Adv. Funct. Mater. 28 (2018).